\documentclass[11pt,a4paper]{article}
\usepackage{jheppub}

\usepackage{color}
\usepackage{amsfonts}
\usepackage{slashed}

\def\tr{{\rm Tr}}

\def\bea{\begin{eqnarray}}
\def\eea{\end{eqnarray}}

\def\lmatrix{\left(\begin{array}}
\def\rmatrix{\end{array}\right)}

\def\msbar{\overline{\rm MS\kern-0.5pt}\kern0.5pt}

\title{The running coupling of 8 flavors and 3 colors}

\author[abc]{Zoltan Fodor,}
\author[d]{Kieran Holland,}
\author[e]{Julius Kuti,}
\author[cf]{Santanu Mondal,}
\author[cf]{Daniel Nogradi}
\author[a]{and Chik Him Wong}

\affiliation[a]{University of Wuppertal, Department of Physics, Wuppertal D-42097, Germany}
\affiliation[b]{J\"ulich Supercomputing Center, Forschungszentrum J\"ulich, J\"ulich D-52425, Germany}
\affiliation[c]{E\"otv\"os University, Institute for Theoretical Physics, Budapest 1117, Hungary}
\affiliation[d]{University of the Pacific, 3601 Pacific Ave, Stockton CA 95211, USA}
\affiliation[e]{University of California, San Diego, 9500 Gilman Drive, La Jolla, CA 92093, USA}
\affiliation[f]{MTA-ELTE Lendulet Lattice Gauge Theory Research Group, 1117 Budapest, Hungary}

\emailAdd{fodor@bodri.elte.hu}
\emailAdd{kholland@pacific.edu}
\emailAdd{jkuti@ucsd.edu}
\emailAdd{santanu@bodri.elte.hu}
\emailAdd{nogradi@bodri.elte.hu}
\emailAdd{rickywong@physics.ucsd.edu}

\abstract{ 
We compute the renormalized running coupling of $SU(3)$ gauge theory coupled to $N_f = 8$ flavors of massless
fundamental Dirac fermions. The recently proposed finite volume gradient flow scheme is used. The calculations are
performed at several lattice spacings allowing for a controlled continuum extrapolation. 
The results for the discrete
$\beta$-function show that it is monotonic without any sign of a fixed point in the range of couplings we cover. As a cross check the continuum
results are compared with the well-known perturbative continuum $\beta$-function for small values of the renormalized coupling
and perfect agreement is found.
}


\arxivnumber{1503.01132}

\begin{document}

\maketitle

\section{Introduction and summary}
\label{intro}

It is well-known that $SU(3)$ gauge theory with $N_f$ flavors of fundamental Dirac fermions has a so-called conformal
window \cite{Caswell:1974gg,Banks:1981nn}. The window refers to the range of flavor numbers where 
the theory possesses an infrared fixed point,
i.e. the long distance behavior is conformal but the short distance behavior is still asymptotically free. Clearly, 
$N_f = 2$ is outside the window and $N_f = 16$ is inside because the perturbative $\beta$-function has a zero at a coupling
where perturbation theory is trustworthy. Unambiguously pinning down the lower end of the conformal
window has been nevertheless difficult so far because only ab initio non-perturbative studies have a chance to settle
the question but these same studies are plagued by systematic uncertainties. It is important to note that these
systematic uncertainties are of a practical nature only and the associated error bars can in principle be reduced to
arbitrary small values. In currently available results with large fermion content the systematic uncertainties were
rarely, if at all, quantitatively estimated. These are however important. The statistical uncertainties can be reduced
by simply increasing the statistics but after a certain point systematic uncertainties will dominate and further increasing the
statistics will be pointless.

The most important of the systematic uncertainties is related to the continuum limit. 
In previous work we have
considered $N_f = 4$ and performed a careful continuum extrapolation; the $N_f = 4$ theory is well-known to be outside the
conformal window and the results obtained were consistent with this expectation. As the next step in approaching the lower end of the
conformal window we continue our investigation with $N_f = 8$ in the present work and pay special attention to
estimating the systematic uncertainties.

The question of finding the lower end of the conformal window is an interesting field theory problem on its own, however there are reasons to be
interested in the $N_f = 8$ model for phenomenological purposes. A large class of Beyond Standard Model extensions 
involve a composite Higgs particle. A natural framework for a composite Higgs state is strong dynamics which solves the
hierarchy problem of the Standard Model and gives rise to dynamical electro-weak symmetry breaking. One challenge
(among others) that all these models should overcome is that 
they should contain a relatively light scalar particle, which when coupled to the Standard Model is to be identified by
the 125 GeV particle found at the LHC in 2012. A light scalar might arise in strongly coupled non-abelian gauge theories
with many fermions if it is not far from the conformal window. Hence the $N_f = 8$ model might play a useful role in
studying the properties of the hypothetical composite Higgs.

Several aspects of the $N_f = 8$ model were studied in the past. The running coupling using the Schroedinger functional
was investigated in \cite{Appelquist:2007hu}, the thermodynamics of the model was studied in \cite{Deuzeman:2008sc} and
hadron spectroscopy was presented in \cite{Fodor:2009wk, Aoki:2014oha}. These studies agreed in their conclusion that
the model is outside the conformal window and spontaneous chiral symmetry breaking takes place. 
In \cite{Cheng:2013eu} the mass anomalous dimension was investigated 
but a definite conclusion whether the model is inside or outside the
conformal window could not be drawn from the data. In any case conformal behavior was not ruled out.
Finally \cite{Hasenfratz:2014rna} studied the running coupling using our finite volume running coupling scheme
\cite{Fodor:2012td, Fodor:2012qh} as in our present work and in section \ref{conclusion} we will
comment on the relationship between our results and the analysis in \cite{Hasenfratz:2014rna}.

The organization of the paper is as follows. In section \ref{thegradientflow} we briefly summarize the finite volume
gradient flow running coupling scheme that we use; for more details see \cite{Fodor:2012td, Fodor:2012qh}. In section
\ref{numericalsimulations} the details of our numerical simulations are given and the collected data is presented, while
in section \ref{continuumextrapolation} the continuum extrapolation of the discrete $\beta$-function is 
performed. Finally in section \ref{conclusion} we end with a
conclusion and provide avenues for future studies.

\section{The gradient flow running coupling scheme}
\label{thegradientflow}

In order to investigate the infrared behavior of a model the running coupling is a natural choice. There are many
well-defined schemes and one is free to choose any one of them. If the $\beta$-function in one scheme has a non-trivial zero
indicating conformal behavior in the infrared, its existence is universal in every other well-defined scheme. In the
current work the recently proposed finite volume gradient flow scheme \cite{Fodor:2012td, Fodor:2012qh} is used, which
is based on Luscher's Wilson flow \cite{Luscher:2009eq, Luscher:2010iy, Luscher:2010we, Luscher:2011bx} 
related to earlier constructions by Morningstar and Peardon \cite{Morningstar:2003gk} as well as 
Lohmayer and Neuberger \cite{Lohmayer:2011si}. In this scheme
a 1-parameter family of couplings is defined in finite 4-volume by 
\bea 
\label{g} 
g_c^2 = \frac{128\pi^2\langle t^2 E(t) \rangle}{3(N^2-1)(1+\delta(c))}\;,\qquad E(t) = -\frac{1}{2} \tr F_{\mu\nu}
F_{\mu\nu}(t)\;,
\eea 
where $N$
corresponds to the gauge group $SU(N)$, $t$ is the flow parameter, $c = \sqrt{8t}/L$ is a constant, $E(t)$ is the field
strength squared at $t$ and 
\bea 
\label{delta} 
\delta(c) = - \frac{c^4 \pi^2}{3} + \vartheta^4\left(e^{-1/c^2}\right) - 1\;,  
\eea 
where $\vartheta$ is the $3^{rd}$ Jacobi elliptic function.
The numerical factors are chosen such that at leading order $g_c^2 = g_{\msbar}^2$ for all $c$. The gauge field is
chosen to be periodic and the massless fermions are anti-periodic in all 4 directions. 
The coupling
$g_c(\mu)$ runs via the scale $\mu = 1/L$; for more details on the gradient flow in general see \cite{Luscher:2009eq,
Luscher:2010iy, Luscher:2010we, Luscher:2011bx} while more details on the finite volume gradient flow scheme can be
found in \cite{Fodor:2012td, Fodor:2012qh}.

A peculiar but well-known property of the femtoworld in a periodic box \cite{Luscher:1982ma,Koller:1985mb,Koller:1987fq,vanBaal:1988va,vanBaal:1988qm,KorthalsAltes:1985tv,Coste:1985mn,Coste:1986cb,KorthalsAltes:1988is}
or small volume dynamics is that the perturbative expansion of $g_c^2$ in
$g_{\msbar}$ contains both even and odd powers, $g_c^2 = g_{\msbar}^2 ( 1 + O(g_{\msbar}) )$ for $N > 2$. This
property results in only the 1-loop $\beta$-function coefficient being the same as the well-known coefficient in
$\msbar$. The case of
$N=2$ has been described in \cite{Fodor:2012qh} and in the present work we focus on $N=3$. 

There are two considerations affecting the choice of the constant $c$ of the 1-parameter family. If chosen too small cut-off effects will be
large, if chosen too large the statistical errors will be large. In \cite{Fodor:2012td, Fodor:2012qh}
we have found for $N_f = 4$ that the value
$c = 3/10$ was optimal in this sense and will use $c = 3/10$ also in the current work. We will however show some
results at $c = 1/5$ in order to illustrate the $c$-dependence of our procedure. Henceforth the index $c$ will
nevertheless be dropped and it will be clear from the context when $c \neq 3/10$.

The expression (\ref{g}) for the coupling was so far considered in the continuum. In a previous work \cite{Fodor:2014cpa} 
we have determined the lattice
spacing dependence of the tree-level correction factor $\delta(c)$. The tree-level, finite volume and finite lattice
spacing perturbative calculation led to the expression
\bea
\label{deltacL}
1 + \delta(c,a/L) &=& \frac{2\pi^2c^4}{3} + \frac{\pi^2 c^4}{3} 
\sum_{n_\mu = 0,\; n^2\neq 0}^{L/a-1} \tr\, \left( e^{-t\left({\cal S}^f + {\cal G}\right)} ({\cal S}^g + {\cal G})^{-1}
e^{-t\left({\cal S}^f + {\cal G}\right)} {\cal S}^e \right), 
\eea
where $p_\mu = 2\pi n_\mu/L$ with an integer non-zero 4-vector $n_\mu$ and
${\cal S}^{f,g,e}(p)$ are the tree-level expressions for the action along the flow, dynamical gauge action and the
observable in momentum space, respectively and ${\cal G}(p)$ is a gauge fixing term. The finite lattice momentum sums
can easily be evaluated numerically to arbitrary precision. See \cite{Fodor:2014cpa} for more details.
This expression can be used to tree-level improve the coupling by simply introducing
\bea
\label{gg}
g^2 = \frac{128\pi^2\langle t^2 E(t) \rangle}{3(N^2-1)(1+\delta(c,a/L))}\;.
\eea
However, we have found that even though tree-level
improvement worked very well in reducing the slope of the continuum
extrapolations with $N_f = 4$ in \cite{Fodor:2014cpa} it only reduced the slope
for small $g^2$ and actually increased it for larger $g^2$ in the current work with $N_f = 8$.
The reason probably lies in the fact that the larger $N_f$ is, the larger the fermion 
loop contributions are. And of course tree-level
improvement is not sensitive to fermionic radiative corrections. We will illustrate these issues in some select cases in section
\ref{continuumextrapolation}.

It should be noted that Schroedinger functional boundary conditions can also be implemented together with the gradient
flow leading to a closely related scheme \cite{Fritzsch:2013je, Fritzsch:2013hda, Ramos:2013gda,
Rantaharju:2013bva, Luscher:2014kea}.
An advantage of the periodic boundary conditions used in the present work is that translational
symmetry is not broken in the time direction. A third option is using twisted boundary conditions which
was explored for $SU(2)$ pure gauge theory to high precision in \cite{Ramos:2014kla}.

The application of the gradient flow is not by any means limited to running
coupling studies, applications also include scale setting in QCD \cite{Borsanyi:2012zs, Borsanyi:2013bia, Bazavov:2013gca, Sommer:2014mea},
thermodynamics \cite{Asakawa:2013laa}, renormalized energy
momentum tensor \cite{Suzuki:2013gza, DelDebbio:2013zaa, Makino:2014taa}, various aspects of chiral symmetry 
\cite{Bar:2013ora, Luscher:2013cpa, Shindler:2013bia} and scalar glueballs \cite{Chowdhury:2014kfa}.

\section{Numerical simulation}
\label{numericalsimulations}

The technical details of the simulations closely follow our work on $N_f = 4$ in \cite{Fodor:2012td, Fodor:2012qh}. 
In particular we use the staggered fermion action
with 4 steps of stout improvement with $\varrho = 0.12$ \cite{Morningstar:2003gk}. The bare fermion mass is
set to zero and anti-periodic boundary conditions in all four directions are imposed on the fermions and the gauge field
is periodic. The gauge action is the tree-level improved Symanzik action \cite{Symanzik:1983dc, Luscher:1984xn}.
The observable $E(t)$ is discretized by the clover-type construction as in \cite{Luscher:2010iy}. 

Along the gradient flow we use two discretizations, the Wilson plaquette action and the tree-level
improved Symanzik gauge action. These setups correspond to the $WSC$ and $SSC$ cases in the terminology of
\cite{Fodor:2014cpa}: the notation is Flow-Action-Observable and $W$ stands for Wilson plaquette action, $S$ for 
tree-level improved Symanzik action and $C$ for the clover discretization.
Both setups lead to the same continuum limit, only the size of cut-off effects is different. This
fact allows for the introduction of yet another coupling definition at finite lattice spacing, which however again leads
to the same continuum limit \cite{alberto},
\bea
g_X^2 = X g_{SSC}^2 + (1-X) g_{WSC}^2\;.
\label{mix}
\eea
Here the parameter $X$ is arbitrary, the choice of the two coefficients, $X$ and $1-X$, guarantees that the continuum
limit of $g_X^2$ is the same as that of $g_{SSC}^2$ or $g_{WSC}^2$, i.e. the correct one. 
It is important to note that $X$ is a constant and
does not depend on the bare gauge coupling $\beta$ or the lattice volume $L/a$. In practice we have found that the
choice $X = 1.75$ is most useful. Note that in principle $X$ could depend on the renormalized coupling but in the
present work we do not explore this possibility.

\begin{table}
\begin{center}
\begin{tabular}{|l||l|l|l|l|l|l|l|l|l|l|l|}
\hline
$L/a\;\;\;\;\beta$&3.5&    3.6 &      3.7 &      4.0 &      4.5 &      5.0  \\
\hline
\hline
12 &   6.42(4)&   5.85(4)&   5.29(2)&   4.00(2)&  2.775(7)&   2.12(1) \\
\hline
16 &   7.66(6)&   6.94(4)&   6.28(2)&   4.67(4)&   3.19(1)&   2.43(2) \\
\hline
18 &   8.17(7)&          &    6.6(1)&   4.95(3)&   3.36(2)&   2.52(1) \\
\hline
20 &   8.55(5)&   7.77(4)&   6.98(3)&   5.17(3)&   3.51(2)&   2.63(1) \\
\hline
24 &   9.33(8)&   8.51(5)&          &   5.50(7)&   3.68(2)&   2.76(2) \\
\hline
30 &   10.4(1)&          &    8.3(1)&   6.03(9)&   4.02(4)&   2.93(3) \\
\hline
36 &          &   10.2(1)&          &   6.52(8)&   4.19(4)&   3.07(4) \\
\hline
\hline
\hline
$L/a\;\;\;\;\beta$&6.0&    7.0 &      8.0 &      9.5 &      15.0& \\
\hline                                                       
\hline                                                       
12 &  1.444(6)&  1.098(4)&  0.890(2)&  0.696(2)&  0.383(2)& \\
\hline                                                       
16 &   1.64(1)&  1.242(6)&  1.000(6)&  0.774(3)&  0.426(1)& \\
\hline                                                       
18 &  1.704(6)&  1.288(4)&  1.035(7)&  0.799(4)&  0.437(1)& \\
\hline                                                       
20 &  1.757(8)&  1.322(5)&  1.062(5)&  0.820(1)&  0.449(2)& \\
\hline                                                       
24 &   1.84(1)&  1.376(7)&  1.099(6)&  0.847(4)&  0.463(2)& \\
\hline                                                       
30 &   1.93(2)&   1.43(1)&  1.141(4)&  0.880(4)&  0.481(3)& \\
\hline                                                       
36 &   2.00(2)&   1.48(2)&   1.17(1)&   0.90(1)&          & \\
\hline
\end{tabular}
\end{center}
\caption{Measured renormalized couplings $g^2(L)$ for given bare couplings $\beta$ and lattice sizes $L/a$ using the
linear combination method with $X=1.75$ at $c=3/10$.}
\label{nf08table}
\end{table}

Just as in \cite{Fodor:2012td, Fodor:2012qh} where $N_f = 4$ was considered we do not need to take the root of the
fermion determinant. Hence the results do not depend on the validity of the fourth-root-trick commonly used for QCD. The
evolution along a trajectory of the hybrid Monte Carlo algorithm \cite{Duane:1987de} is implemented
with multiple time scales \cite{Sexton:1992nu} and Omelyan integrator \cite{Takaishi:2005tz}.

In a lattice setting the most practical method of calculating the running coupling or its $\beta$-function is via step
scaling \cite{Luscher:1991wu, Luscher:1992an}. In this context the linear size $L$ is increased by a factor $s$
and the difference of couplings
\bea
\label{discr}
\frac{ g^2(sL) - g^2(L)}{\log( s^2 ) }\;,
\eea
is defined as the discrete $\beta$-function.
If the ordinary infinitesimal $\beta$-function of the theory possesses an infrared fixed point, 
the discrete $\beta$-function will
have a zero as well. On the lattice the linear size $L$ is easily increased to $sL$ by simply increasing
the volume in lattice units, $L/a \to sL/a$ at fixed bare gauge coupling. In the current work we set $s = 3/2$ and use
volumes $12^4 \to 18^4,\; 16^4 \to 24^4,\; 20^4 \to 30^4$ and $24^4 \to 36^4$. 
The continuum limit corresponds to $L/a \to \infty$. These lattice volumes determine the $\beta$-function at 
4 lattice spacings, allowing for a fully controlled continuum
extrapolation. Leading cut-off effects are known to be $O(a^2/L^2)$.

\begin{figure}
\begin{center}
\includegraphics[width=14cm]{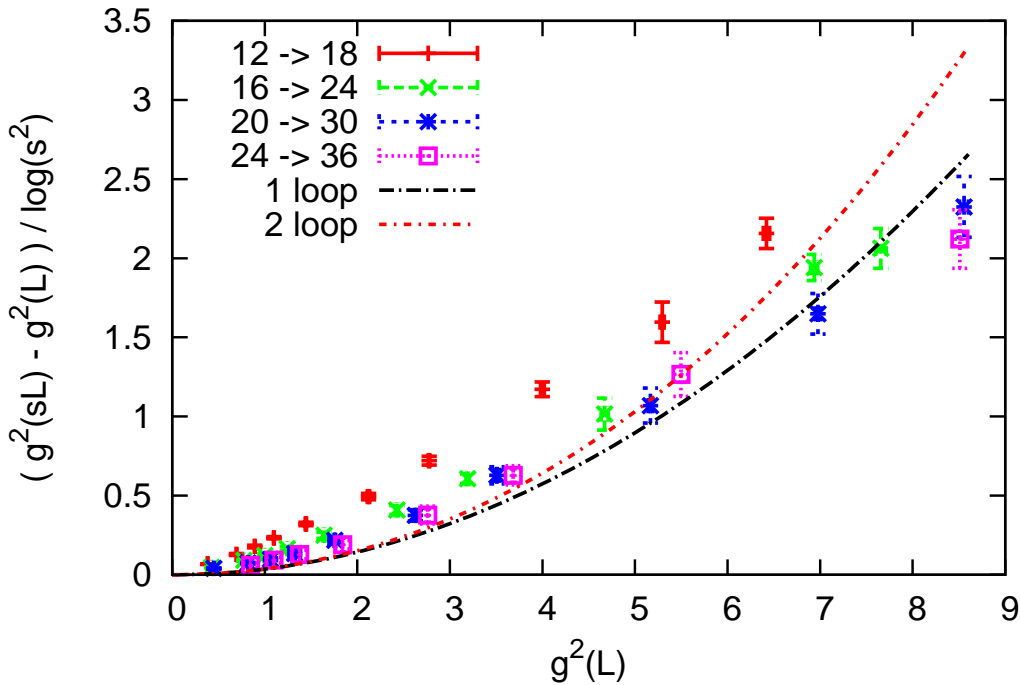}
\end{center}
\caption{Measured discrete $\beta$-function for the linear combination setup with $X=1.75$ and $c=3/10$; data corresponding to four lattice spacings.}
\label{beta}
\end{figure}

The collected number of thermalized trajectories at each bare coupling and volume was in the range between 5000 and 10000 and every
10$^{th}$ was used for measurement. The measured renormalized couplings at each $\beta$ and lattice volume are
shown in table \ref{nf08table} for the definition (\ref{mix}) using $X=1.75$.
By taking the difference of renormalized couplings for lattice volumes scaled by a factor $s=3/2$
and at the same bare $\beta$ one obtains the discrete $\beta$-function at finite lattice spacings; see figure
\ref{beta}. Clearly, there is no sign of a fixed point, the running is monotonically increasing, 
at least at finite lattice spacing, i.e. finite lattice volumes. 
However we are of course interested in the behavior of the continuum model and the behavior of the discrete
$\beta$-function on finite lattice volumes is irrelevant. It is a priori possible that the discrete $\beta$-functions on several 
finite lattice volumes, corresponding to a fixed set of $L/a \to sL/a$ steps, cross zero but the continuum
extrapolated result does not have a zero and conversely it is possible that none of the finite lattice volume
$\beta$-functions cross zero yet the continuum extrapolated result does have a zero.
Hence we turn to the continuum extrapolation next.

\section{Continuum extrapolation}
\label{continuumextrapolation}

In order to perform a continuum extrapolation we parametrize the renormalized coupling as a function of the bare
coupling, $g^2(\beta)$ at each fixed lattice volume $L/a$ by
\bea
\label{p}
\frac{1}{g^2(\beta)} = \frac{\beta}{6} \sum_{m=0}^n C_m \left( \frac{6}{\beta} \right)^m\;,
\eea
similarly as in \cite{Tekin:2010mm}. The order $n$ of the polynomial may be chosen such that acceptable fits are
obtained, however in this work we would like to estimate the systematic errors that come from various choices for $n$;
see section \ref{systematicerror}.

\begin{figure}
\begin{center}
\includegraphics[width=8.5cm]{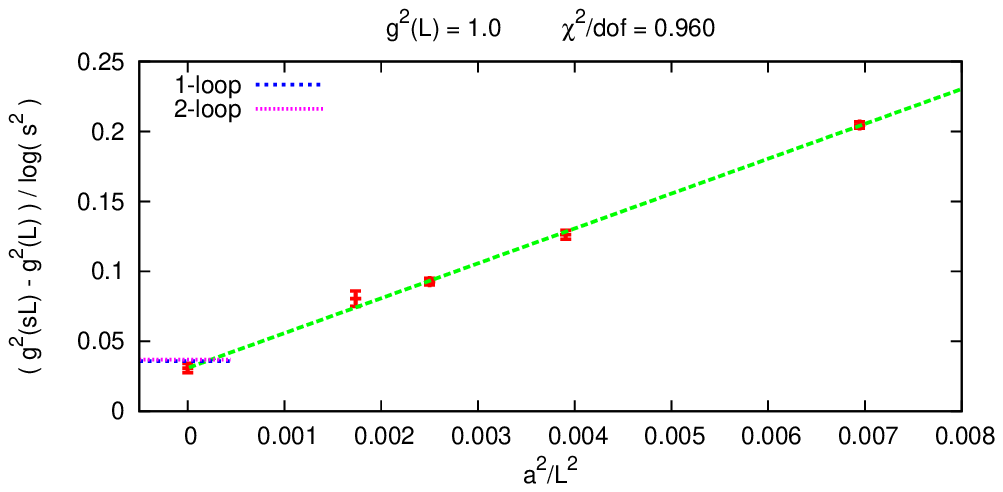} \includegraphics[width=6.4cm]{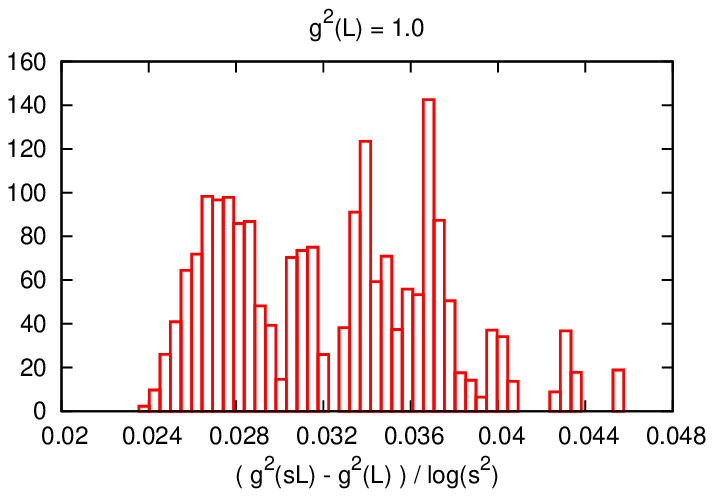} \\
\includegraphics[width=8.5cm]{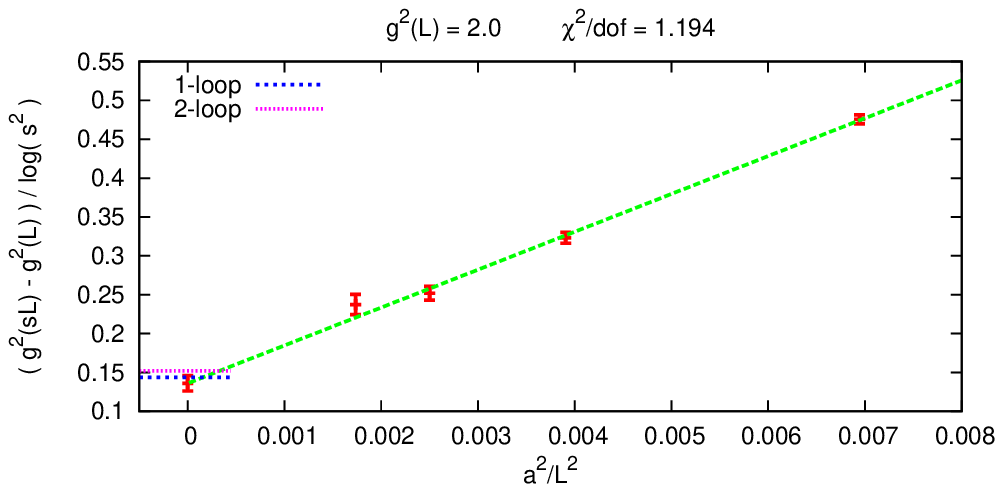} \includegraphics[width=6.4cm]{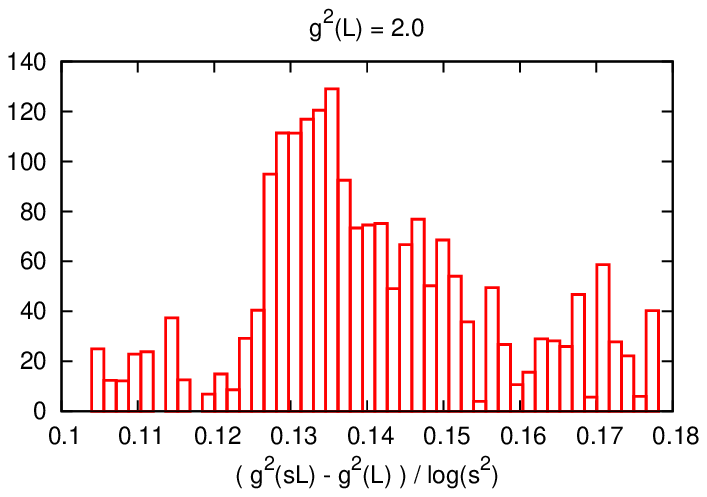} \\
\includegraphics[width=8.5cm]{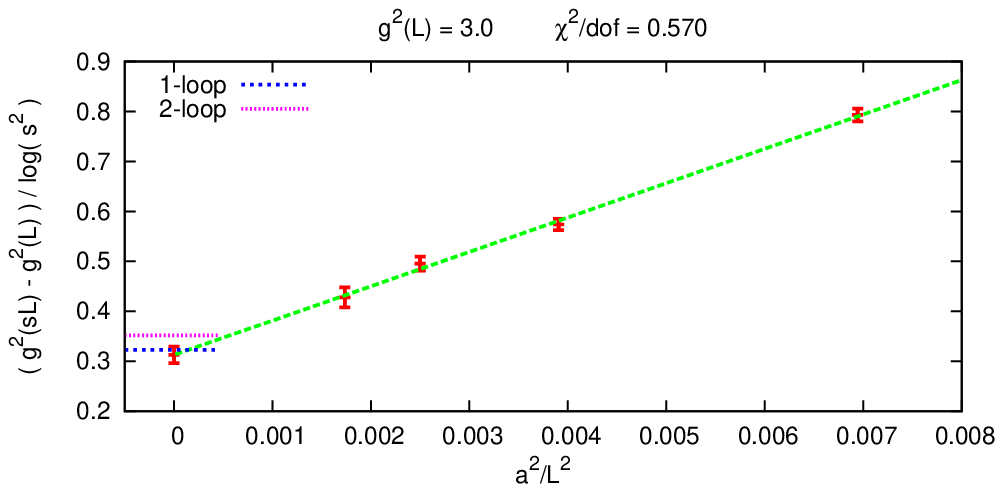} \includegraphics[width=6.4cm]{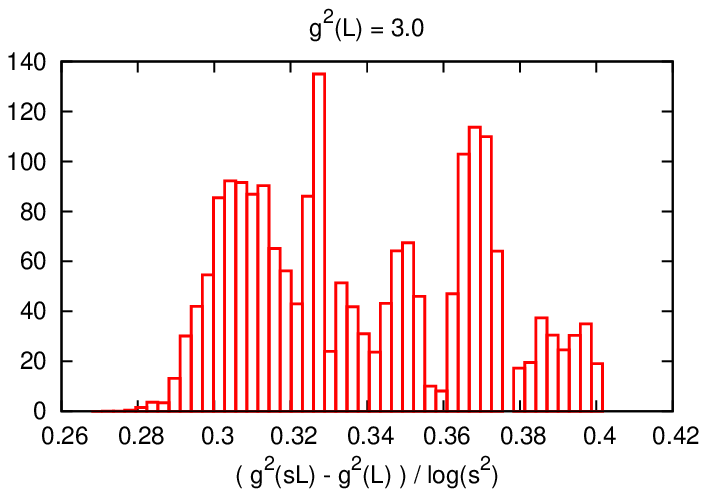} \\
\end{center}
\caption{Right: the weighted histograms of all possible continuum extrapolations used for estimating the systematic
uncertainty. Left: a representative example of the continuum extrapolations for $g^2(L) = 1.0, 2.0, 3.0$; the 1-loop 
and 2-loop results are also shown for comparison. All data is
with $c=3/10$ and using the linear combination method with $X=1.75$.}
\label{somegsq1}
\end{figure}

\begin{figure}
\begin{center}
\includegraphics[width=8.5cm]{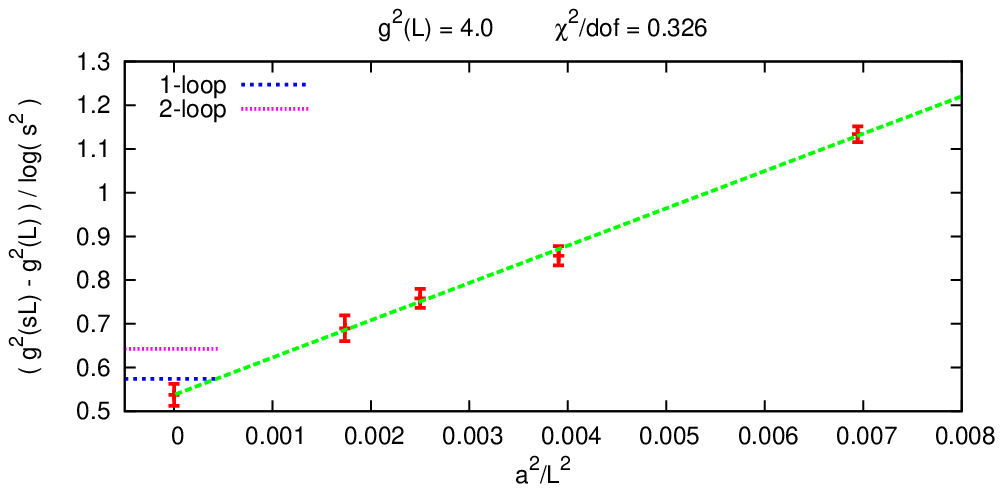} \includegraphics[width=6.4cm]{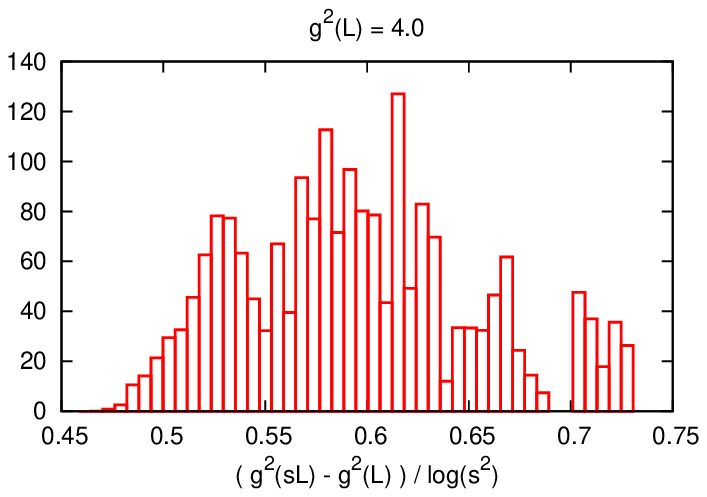} \\
\includegraphics[width=8.5cm]{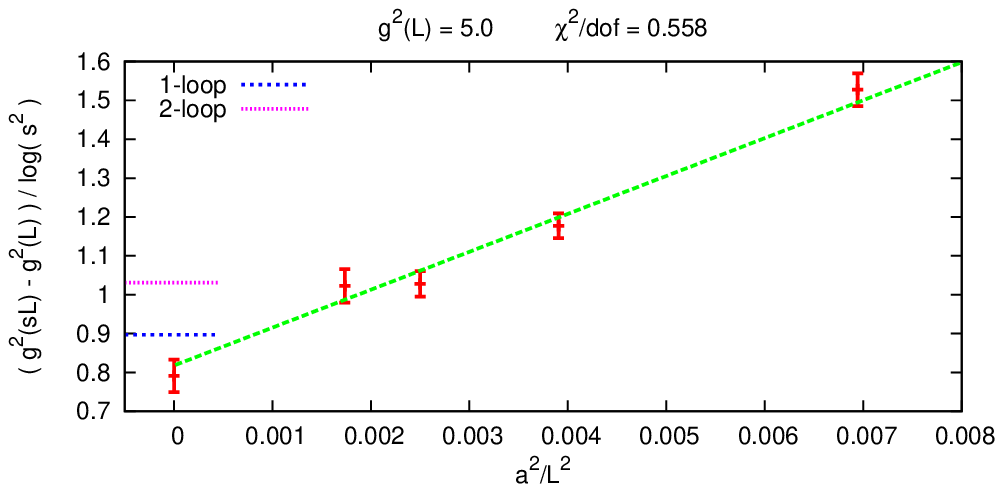} \includegraphics[width=6.4cm]{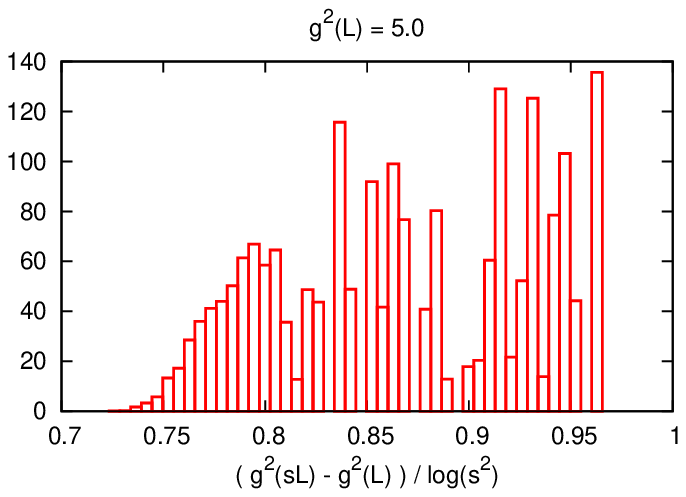} \\
\includegraphics[width=8.5cm]{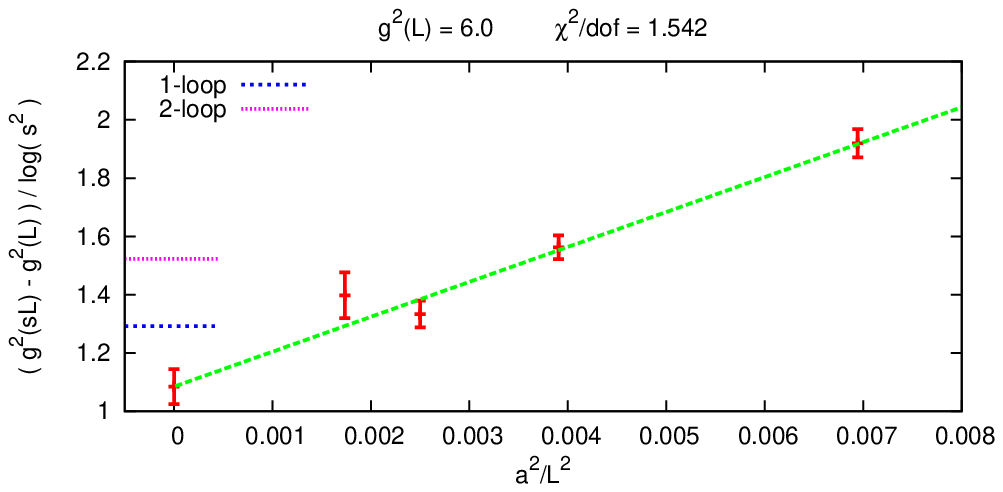} \includegraphics[width=6.4cm]{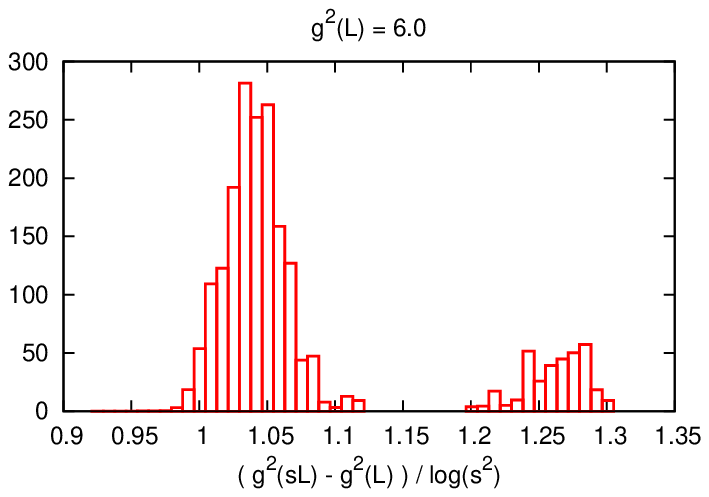} \\
\end{center}
\caption{Right: the weighted histograms of all possible continuum extrapolations used for estimating the systematic
uncertainty. Left: a representative example of the continuum extrapolations for $g^2(L) = 4.0, 5.0, 6.0$; the 1-loop 
and 2-loop results are also shown for comparison. All data is
with $c=3/10$ and using the linear combination method with $X=1.75$.}
\label{somegsq2}
\end{figure}

Using the parametrized curves the discrete $\beta$-function (\ref{discr}) can be obtained for
arbitrary $g^2(L)$ for fixed $L/a$ and $s = 3/2$. Estimating the error on the interpolated values is straightforward
because the interpolation is linear in the fit parameters $C_m$. Then assuming that corrections are linear 
in $a^2/L^2$ the continuum extrapolation can be performed.

\subsection{Systematic error}
\label{systematicerror}

In our previous work \cite{Fodor:2012td, Fodor:2012qh} the polynomial order for the interpolation (\ref{p}) was fixed at
each lattice volume. However different choices lead to similarly acceptable interpolating fits and these in turn lead to
slightly different continuum results. Even though the final continuum result varies only a bit and generally within
1-$\sigma$ of the statistical error in the current work we would like to estimate the systematic error as precisely as
possible. In order to achieve this the histogram method introduced in \cite{Durr:2008zz} is used. There are two sources
of systematic uncertainties. First, it is a priori unknown what interpolation function to use for the renormalized
coupling as a function of $\beta$ at fixed lattice volumes, and second, one may perform continuum extrapolations using 3
or 4 lattice spacings.

We interpolate using (\ref{p}) for each lattice volume, $12^4$, $16^4$, $18^4$, $20^4$, $24^4$, $30^4$, $36^4$, with
three choices of polynomial orders, $n=4,5$ and $6$. All together these produce $3^7 = 2187$ combination of
interpolations and correspondingly
lead to $2187$ different continuum results. Since the data on different volumes at different $\beta$ are all independent
we perform a Kolmogorov-Smirnov test on the $2187$ interpolations and demand that only those assignments of polynomial
orders are allowed to which the Kolmogorov-Smirnov test assigns at least a $30\%$ probability, similarly to
\cite{Borsanyi:2014jba}. 

The Kolmogorov-Smirnov test is applied as follows \cite{Borsanyi:2014jba}. 
The $\chi^2$ values of independent fits are distributed according to
the $\chi^2$-distribution. The goodness of fits, or q-values, are on the other hand distributed uniformly. The
Kolmogorov-Smirnov test is an estimate of the probability that the actual measured q-values were indeed distributed
uniformly. The cumulative distribution function of the uniform distribution is a straight line and the
Kolmogorov-Smirnov test takes as input the largest distance between the actual measured cumulative distribution 
function and the expected cumulative distribution function (straight line). Call this largest distance $D$. Then the
Kolmogorov-Smirnov probability is defined by
\bea
P = Q\left( D \left( \sqrt{ N } + 0.12 + \frac{ 0.11 }{ \sqrt{ N } } \right) \right)\;,\qquad 
Q(x) = 1 - \vartheta_4\left(e^{-2x^2}\right)
\eea
where $\vartheta_4$ is the $4^{th}$ Jacobi elliptic function and $N$ is the sample size \cite{nr}.

The Kolmogorov-Smirnov test with $P>0.3$ reduced
the total number of allowed interpolations from $2187$ to $1233$ as far as 4 lattice spacings are concerned corresponding to
$12^4 \to 18^4$, $16^4 \to 24^4$, $20^4 \to 30^4$ and $24^4 \to 36^4$.

In order to include the systematic uncertainty from the continuum {\em extrapolation} itself, as opposed to the {\em
interpolation} at fixed lattice volume, we consider dropping the roughest lattice spacing corresponding to $12^4 \to
18^4$ and use only $16^4 \to 24^4$, $20^4\to 30^4$ and $24^4 \to 36^4$. From the $1233$ continuum extrapolations using 4
lattice spacings only those extrapolations using 3 lattice spacings are kept to which again the Kolmogorov-Smirnov test
assigns a probability larger than $30\%$, in terms of the 5 independent volumes, $16^4, 20^4, 24^4, 30^4, 36^4$. This
test leads to $813$ continuum extrapolations using 3 lattice spacings. Some of these are of course the same, but needs to
be counted in order to have the proper weight in the final histogram.

The $1233 + 813 = 2046$ continuum results at each $g^2(L)$ can be binned in a weighted histogram 
and the weight can be the goodness of the fit, a
weight provided by the Akaike information criterion (AIC) \cite{aic1, aic2, aic3} or no weight at all. 
If a fit has $p$ free parameters its associated AIC weight is $\sim \exp(-\chi^2/2 - p)$.
Examples of AIC-weighted histograms are shown in figures \ref{somegsq1}-\ref{somegsq2}.

Our continuum central values at each $g^2(L)$ are the medians of the histograms and the systematic uncertainty
can then be determined by counting $68\%$ of the total starting symmetrically from the central value. 
The three types of weights lead
to compatible results and for our final results we use the AIC-weighted histograms.

The systematic and statistical errors are of the same order, there is never a larger factor between them than two.

\subsection{Final results}
\label{finalresults}

At 6 chosen values of $g^2(L)$ the histograms of the discrete $\beta$-function for all continuum extrapolations are
shown in the right panels of figures \ref{somegsq1}-\ref{somegsq2}. On the left we show typical continuum extrapolations
from within a $1-\sigma$ systematic uncertainty around the median of the histograms.
Clearly, all
4 lattice spacings are in the scaling region and nicely fit on a straight line with good $\chi^2/dof$. In fact, the choice $X=1.75$ was
motivated by exactly the requirement that all 4 lattice spacings should be in the scaling region. This is not a sharp
requirement, one may choose any value in the approximate range $1.6 < X < 1.9$.

It is quite instructive to look at the details of these figures and discuss
the source of the most important systematic error, the continuum
extrapolation. Our theory is a confining one in which large bare couplings
(small $\beta$s) correspond to large lattice spacings. As table \ref{nf08table} shows
large renormalized couplings are obtained with large lattice volumes and
small $\beta$ values. Thus, for a given renormalized coupling one reaches the
continuum limit by increasing both $\beta$ and the lattice volume. Since the
largest volume, independently of $\beta$, was $36^4$, large renormalized
couplings correspond within our parameter set to large lattice spacings and
obviously large cutoff effects.

It is of obvious interest to turn this qualitative statement to a
quantitative one and to determine the size of the systematic uncertainty
related to this question. Most importantly, we want to know where to stop with
the present lattice sizes because no controlled continuum extrapolation can
be carried out any further. As our $g^2=6$ case illustrates for this large value of the
renormalized coupling one has a two peak structure for the histogram. The two
peaks are the result of the significant difference between using only the
finer lattices with 3 points or taking 4 points (including also the coarsest
lattices) for the continuum extrapolations. This phenomenon clearly indicates
that the results from the coarsest lattices are starting to deviate from the
$a^2$ scaling showed by the finer lattices. The difference between the peaks
still quantifies the systematic uncertainty for $g^2=6$ and tells us that for
even larger $g^2$ values the control over this systematic effect could be
lost and finer lattices with larger lattice volumes are needed.

The discrete $\beta$-function may reliably be calculated in (continuum) perturbation theory for small values of the
renormalized coupling. In terms of the well-known infinitesimal 1 and 2 loop $\beta$-function coefficients, 
$b_1$ and $b_2$ the discrete variant is given by
\bea
\label{dbeta}
\frac{g^2(sL)-g^2(L)}{\log(s^2)} = b_1 \frac{g^4(L)}{16\pi^2} + \left( b_1^2 \log(s^2) + b_2 \right)
\frac{g^6(L)}{(16\pi^2)^2} + \ldots 
\eea
As noted already in our finite volume gradient flow scheme only $b_1$ is the same as in every other well-defined scheme.
The reason is a well-understood feature of the finite 4-volume or femtoworld \cite{Fodor:2012td, Fodor:2012qh}.
Nevertheless we include not only the 1-loop continuum $\beta$-function but also the 2-loop 
approximation in our comparisons, even though strictly speaking agreement is only expected with the 1-loop
result.

\begin{figure}
\begin{center}
\includegraphics[width=10cm]{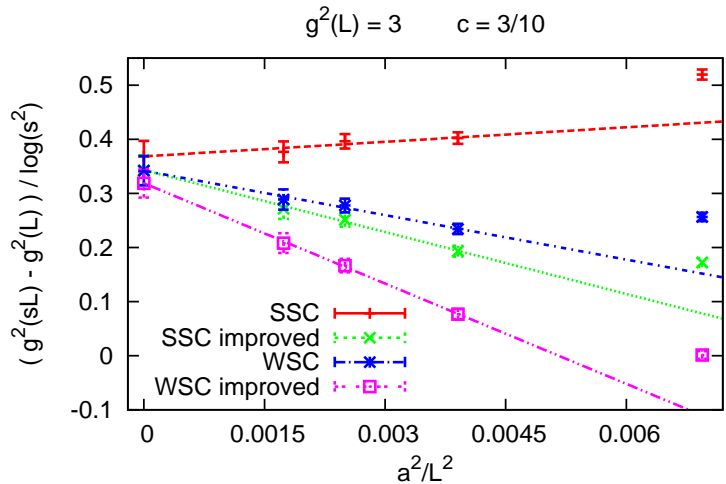}
\end{center}
\caption{Comparison of the tree-level improved and unimproved continuum extrapolations for the $SSC$ and $WSC$ cases
at $c=3/10$. 
Clearly the roughest lattice spacing corresponding to $12^4 \to 18^4$ is not in the scaling region. The
choice $X=1.75$ does bring this point also into the scaling region however; see text for details.
}
\label{comp}
\end{figure}

Had we not used the linear combination (\ref{mix}) only 3 lattice spacings would have been in the scaling region,
$16^4\to24^4$, $20^4\to30^4$ and $24^4\to36^4$ assuming a fit linear in $O(a^2/L^2)$. As mentioned in section
\ref{numericalsimulations} tree-level improvement \cite{Fodor:2014cpa} did not reduce the slope of the continuum
extrapolations as dramatically as for $N_f = 4$ in our previous study. The reason is presumably that the larger fermion
content results in larger fermionic contributions which are, of course, completely absent from the tree-level expressions. We
illustrate both points, the smaller scaling region without employing the linear combination (\ref{mix}) and the less
effective tree-level improvement in figure \ref{comp}. Clearly, the continuum results are always consistent, as they
should be, the various choices (improvement vs. non-improvement, linear combination vs. no linear combination) only
affect the slopes of the extrapolations and the size of the scaling region.

\begin{figure}
\begin{center}
\includegraphics[width=7.5cm]{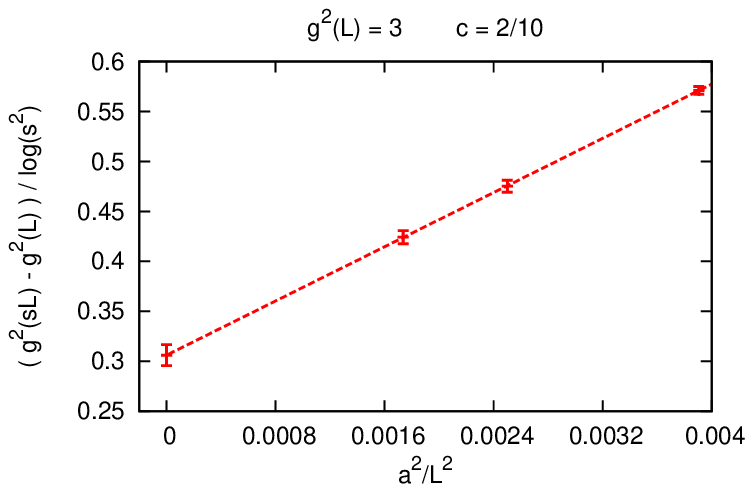} \hspace{-0.5cm} \includegraphics[width=7.5cm]{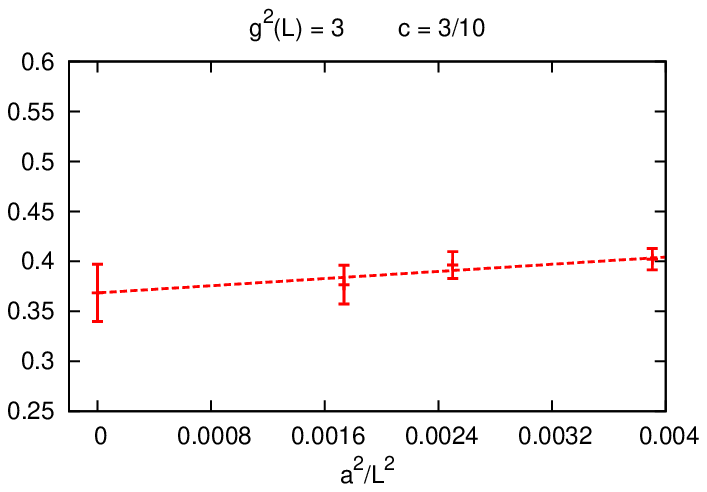}
\end{center}
\caption{Continuum limit of the (unimproved) $SSC$ setup at $g^2(L)=3.0$; comparing $c=2/10$ (left) and $c=3/10$
(right). Clearly, as $c$ increases the cut-off effects become smaller but the statistical errors grow. The continuum
extrapolations do not have to agree since different $c$ values correspond to different schemes.}
\label{cdep}
\end{figure}

In figure \ref{cdep} we illustrate another aspect mentioned in section \ref{numericalsimulations}, namely
$c$-dependence. Different choices of $c$ define different schemes, i.e. the $\beta$-function will be $c$-dependent. For
small coupling the difference should be very small since regardless of what $c$ is, agreement is expected with the
perturbative 1-loop result. Furthermore, the expectation is that a smaller $c$ leads to smaller 
statistical errors because of smaller
autocorrelations and also to larger cut-off effects because of the smaller flow time $t$. This is illustrated in figure
\ref{cdep} where the continuum extrapolation is shown for $g^2 = 3$ and both for $c=3/10$ and $c=1/5$.

\begin{figure}
\begin{center}
\includegraphics[width=10cm]{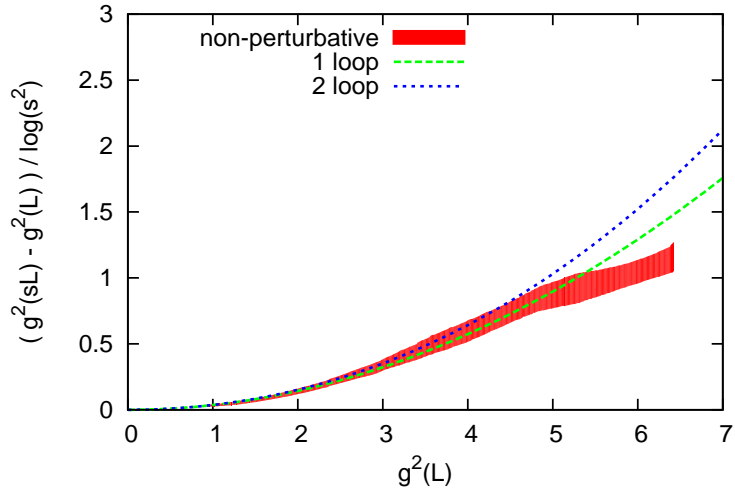}
\end{center}
\caption{Our final result for the continuum extrapolated discrete $\beta$-function.}
\label{cont}
\end{figure}

Finally, in figure \ref{cont} we show the continuum extrapolated $\beta$-function over the entire $0.9 < g^2 < 6.3$ range
accessible to our simulations together with the 1-loop and 2-loop results. The linear combination method (\ref{mix}) was
used with $X=1.75$ and $c=3/10$ was chosen. 
Our non-perturbative continuum result is in nice agreement with the 
perturbative results for small renormalized coupling and deviates from it
for larger values. Most importantly, the deviation from the perturbative 1-loop result is downward. This could have been expected
because at some higher $N_f$ value we do expect a fixed point and by continuity one might argue that this is only
possible if the running is slower than the monotonically increasing 1-loop result, at least for some $N_f$ value which is
not far below the conformal window. At $N_f = 8$ we do not see a sign of a fixed point in any case, at least in the
explored range $0.9 < g^2 < 6.3$.

\section{Conclusion and outlook}
\label{conclusion}

In this work we have continued our study of $SU(3)$ gauge theory with many fermions. The representation was fundamental
and after having examined $N_f = 4$ in our previous work the $\beta$-function of the $N_f = 8$ model was computed in the present work, in
the continuum. The $\beta$-function does not appear to ``bend back'' in the coupling range we have studied hence does
not support the idea that the $N_f = 8$ model is already inside the conformal window. 
This result is consistent with our study of the
mass spectrum which indicated spontaneous breaking of chiral symmetry at
zero fermion mass \cite{Fodor:2009wk}.
The running coupling does deviate
from the perturbative $\beta$-function downwards though.

While preparing our manuscript the work \cite{Hasenfratz:2014rna} appeared. The method used there is similar to ours and
the conclusions were also similar, i.e. the behavior was compatible with a monotonically increasing $\beta$-function.
The explored coupling range was larger, $2.0 < g^2 < 14$, but the continuum results were not compatible with
weak coupling perturbation theory even at the weakest coupling $g^2 = 2$. 
This is a puzzling feature since one would expect perturbation theory to be
reliable at such a small coupling. 
At $g^2 = 2$ the difference between the 1-loop and 4-loop (in $\msbar$) result is about 1\% suggesting that perturbation
theory is indeed trustworthy (we expect the odd terms in the gauge coupling to be small as well). The difference between the
1-loop and the continuum extrapolated result of \cite{Hasenfratz:2014rna} is however around 40\%. 
\footnote{We would like to thank the referee for pointing out that the
discrepancy is $2.5\,\sigma$ between the lattice result \cite{Hasenfratz:2014rna} and perturbation theory in a finite range of
the weak coupling regime.}

In our work we in fact show consistency with perturbation theory up to approximately $g^2 =
5$ and only detect deviations for larger couplings which is more in line with expectations. The reason for the
discrepancy in \cite{Hasenfratz:2014rna} might be due to the fact that the systematic uncertainties were not adequately
addressed. In our work we controlled both types, one from the a priori unknown interpolation as a function of the bare
coupling at finite lattice volume and also the one coming from the continuum extrapolation.

\acknowledgments

This work was supported by the DOE  grant DE-SC0009919,
by the Deutsche Forschungsgemeinschaft
grants FO 502/2 and SFB-TR 55 and by the NSF under grants 0704171, 0970137 and 1318220 
and by OTKA under the grant OTKA-NF-104034.
Computations were carried out at the PC clusters of Fermilab, 
on the GPU clusters at the University of Wuppertal, on Juqueen at FZJ and the Eotvos
University in Budapest, Hungary and at the University of California San Diego, USA using the 
CUDA port of the code \cite{Egri:2006zm}. We also benefitted from an ALCC Award on the BG/Q Mira
platform of ALCF. Kalman Szabo and Sandor Katz are gratefully acknowledged for code development.

\end{document}